\documentclass[submission]{eptcs}

\usepackage{graphicx}
\usepackage{listings}
\usepackage{caption}
\usepackage{subcaption}
\usepackage[fleqn]{amsmath}
\usepackage{amssymb}
\usepackage{color}
\usepackage{multirow}
\usepackage{xspace}
\usepackage{hyperref}

\definecolor{comment}{rgb}{0.2, 0.6, 0.4}
\definecolor{string}{rgb}{0, 0, 1}
\definecolor{keyword}{rgb}{0.6, 0.2, 0.4}
\definecolor{background}{gray}{0.97}

\providecommand{\event}{WLP 2016}

\newcommand{\astlog}{\textsc{Astlog}\xspace}
\newcommand{\astr}{\textsc{Ast}\xspace}
\newcommand{\astrs}{\textsc{Ast}s\xspace}
\newcommand{\ctl}{\textsc{Ctl}\xspace}

\newcommand{\cg}{\textsc{Cg}\xspace}
\newcommand{\cgs}{\textsc{Cg}s\xspace}
\newcommand{\est}{\textsc{Est}\xspace}
\newcommand{\ests}{\textsc{Est}s\xspace}
\newcommand{\ide}{\textsc{Ide}\xspace}
\newcommand{\ides}{\textsc{Ide}s\xspace}
\newcommand{\llvm}{\textsc{Llvm}\xspace}
\newcommand{\nest}{\textsc{Nest}\xspace}
\newcommand{\nests}{\textsc{Nest}s\xspace}
\newcommand{\rodos}{\textsc{Rodos}\xspace}
\newcommand{\scml}{\textsc{Scml}\xspace}
\newcommand{\sql}{\textsc{Sql}\xspace}
\newcommand{\xml}{\textsc{Xml}\xspace}

\title{Source Code Verification for Embedded Systems using Prolog}

\author{
Frank Flederer
\institute{University of Wuerzburg\\Aerospace Information Technology}
\email{frank.flederer@uni-wuerzburg.de}
\and
Ludwig Ostermayer
\institute{University of Wuerzburg\\Knowledge-based Systems}
\email{ludwig.ostermayer@uni-wuerzburg.de}
\and
Dietmar Seipel
\institute{University of Wuerzburg\\Knowledge-based Systems}
\email{dietmar.seipel@uni-wuerzburg.de}
\and
Sergio Montenegro
\institute{University of Wuerzburg\\Aerospace Information Technology}
\email{sergio.montenegro@uni-wuerzburg.de}
}

\def\titlerunning{%
   Source Code Verification for Embedded Systems using Prolog}
\def\authorrunning{%
   F. Flederer, L. Ostermayer, D. Seipel \& S. Montenegro}

\begin{document}

\nocite{*}

\maketitle

\begin{abstract}
System relevant embedded software needs to be reliable and, therefore, well
tested, especially for aerospace systems.
A common technique to verify programs is the analysis of their abstract syntax
tree (\astr).
Tree structures can be elegantly analyzed with the logic programming
language Prolog. 
Moreover, Prolog offers further advantages for a thorough analysis:
On the one hand, it natively provides versatile options to efficiently process
tree or graph data structures.
On the other hand, Prolog's non-determinism and backtracking eases tests of different variations of the program flow without big effort.
A rule-based approach with Prolog allows to characterize the verification goals
in a concise and declarative way.

In this paper, we describe our approach to verify the source code of a flash
file system with the help of Prolog.
The flash file system is written in C++ and has been developed particularly
for the use in satellites.
We transform a given abstract syntax tree of C++ source code
into Prolog facts and derive the call graph and the execution sequence (tree),
which then are further tested against verification goals.
The different program flow branching due to control structures is derived by backtracking as subtrees of the full execution sequence.
Finally, these subtrees are verified in Prolog.

We illustrate our approach with a case study, where we search for incorrect
applications of semaphores in embedded software using the real-time operating system \rodos.
We rely on computation tree logic (\ctl) and have designed an embedded domain specific language (\textsc{Dsl}) in Prolog
to express the verification goals.
\end{abstract}

\section{Introduction}
\label{sec introduction}

Embedded systems are heavily used in mission critical parts, e.g.\ in the
field of aerospace.
Therefore, established and time-proven software is preferred to modern cutting-edge technology.
Thus, new software concepts must be well tested and approved before it is applied in critical missions.

The majority of software in embedded systems is written in the two programming languages C~(51~\%) and C++~(30~\%)~\cite{Nah:09}; both are often processed by the same tools such as Clang/\llvm.
In a satellite project, we have developed a tailored file system for the embedded use on a satellite.
It makes use of several flash memory chips and one \textsc{Mram}, a non-volatile random access memory.
The file system is a module for the real-time operating system \rodos\footnote{Real-time On-board Dependable Operating System:
http://www.dlr.de/irs/desktopdefault.aspx/tabid-5976/9736\_read-19576/}, which
is developed at the German Aerospace Center (\textsc{Dlr}) and the chair for aerospace
information system at the university of Wuerzburg.
Like \rodos, we have written the file system in C++.
To improve the correct implementation, we want to verify the file system formally.
In extend to general-purpose static analyses of software, we plan to test domain specific patterns for the file system.

\paragraph{Verification Goals}
\label{sec verification goals}
For our first approach of verification, in this paper, we pick one aspect of using operating systems: the application of semaphores.
Nonetheless, our approach can be adapted to test for other verification goals.
\rodos provides semaphores as classes with the methods \verb|enter| and \verb|leave|.
Both methods have no return values (\verb|void|).
However, the method \verb|enter| blocks if the semaphore has been entered previously.
Usually, a programmer uses semaphores by directly accessing its instance variable.
Semaphores are used for synchronizing concurrent threads in \rodos; as long as a semaphore is entered and not left no thread can enter it a second time.
\begin{quote}
\begin{lstlisting}[language=c++]
void SemaTest::methodA() {
   sema.enter();
   methodB();
   sema.leave();
}
\end{lstlisting}
\end{quote}
The semaphore \verb|sema| is entered once and left, eventually.
However, semaphores can be used wrongly.
Entering a semaphore without leaving it is an example for an incorrect usage.
\begin{quote}
\begin{lstlisting}[language=c++]
void Test::methodC() {
   methodE();
   sema.leave();
}
void Test::methodD() {
   methodE();
}
void Test::methodE() {
   methodF();
   sema.enter();
   methodG();
}
void Test::methodF() { }
void Test::methodG() { }
\end{lstlisting}
\end{quote}
We start with the invocation of \verb|methodC|.
It first calls \verb|methodE| which enters the semaphore.
Then, \verb|methodE| is left while keeping the semaphore entered.
After returning to \verb|methodC| the semaphore is left.
Therefore, for \verb|methodC|, the semaphore is used correctly.

However, if \verb|methodD| is invoked, it only calls \verb|methodE| which enters the semaphore.
After that, the execution returns from \verb|methodE| to \verb|methodD|.
However, there are no other calls within \verb|methodD|.
Thus, if the semaphore will not be left after returning from \verb|methodD|, then there is an incorrect usage of semaphores.
Consequently, we have to analyze the sequence of method calls to check the correct application of semaphores.

\paragraph{Overview of the Verification Process}

\begin{figure*}[ht]
\centering
\includegraphics[width=.7\textwidth]{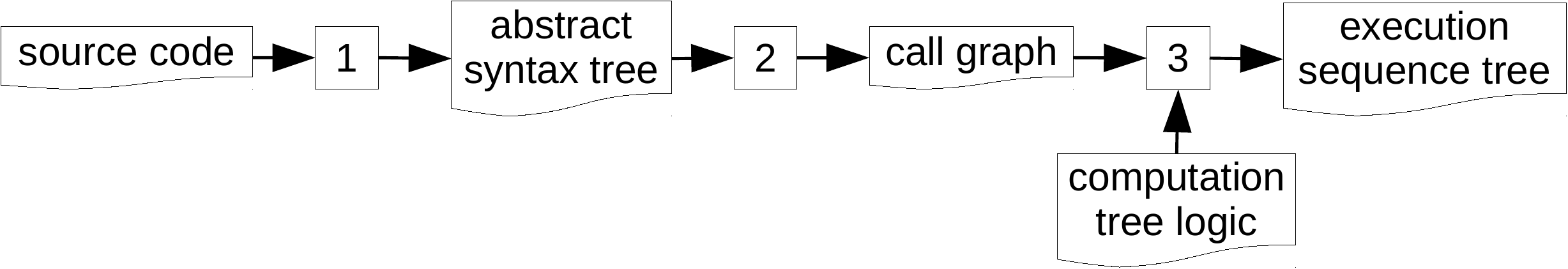}
\caption{The Verification Process is Partitioned into Three Subsequent Stages}
\label{fig process}
\end{figure*}

The basis for our investigation is the source code of the software; we transform it into different graph representations in subsequent stages.
Figure~\ref{fig process} outlines the stages and graphs.
At first, a converter using Clang transforms the corresponding \emph{abstract syntax tree} (\astr) into Prolog facts.
Secondly, declarative Prolog rules match graph patterns of method calls within the \astr and extract information about method calls into a \emph{call graph} (\cg).
In the last stage, traversing Prolog rules construct \emph{execution sequence trees} (\est).
\ests contain information about method calls that are sequentially executed after a given method call.
Based on user-defined \emph{computation tree logic} (\ctl) propositions and different program flows in the software, the process generates different non-deterministic variations of the \est, which we call \emph{non-deterministic execution sequence trees} (\nest).
The traversing process simultaneously evaluates the \ctl propositions on the nodes during the construction of the \nest; therefore, it detects \nests with failing or successful \ctl checks early and, thus, aborts the creation prematurely.
This reduces the number of \nests that are investigated eventually.

\paragraph{Using Prolog for Verification}

Similar to JTransformer\cite{Kni:07}, which operates with Prolog on Java source code, we want to perform static source code analyses by using Prolog rules~-- but for C/C++ software.
This paper shows our first attempt of source code verification for embedded software using non-deterministic Prolog rules.
We use Prolog for several reasons.
Firstly, we are able to extend the analysis by writing comprehensible Prolog rules.
Secondly, Prolog enables an easy definition of domain specific languages to match the requirements of specific software parts, e.g.\ file systems.
Thirdly, as we heavily operate on graph structures, we can use existing graph investigation techniques and libraries which are already available in Prolog systems.
Additionally, we use non-determinism in Prolog.
Control structures in source code lead to different branches of the program flow at runtime, depending on their evaluation results.
We use Prolog's backtracking mechanism to test several variants of program flows for the verification goals.
For our implementation we use the established Prolog system \textsc{Swi}-Prolog \cite{Wie:03}.

\paragraph*{Organization of the Paper}
The remainder of this paper is organized as follows.
Section~\ref{sec ast} describes the representation of the source code as abstract syntax tree.
From the abstract syntax tree we extract in Section~\ref{sec cg} a call graph for methods.
For the investigation of subsequent calls, we create in Section~\ref{sec est} an execution sequence tree.
Based on the execution sequence tree, the control structures in the source code and the verification goals, we generate different non-deterministic execution sequence trees.
In Section~\ref{sec case study} we define appropriate computation tree logic propositions to analyze the usage of semaphores.
Section~\ref{sec related work} summarizes other approaches that use logic programming for software verification.
Finally, we conclude in Section~\ref{sec conclusion} with an outlook to future work.

\section{Representation of the Abstract Syntax Tree in Prolog}
\label{sec ast}

At the first stage (stage 1 in Figure~\ref{fig process}) a converter translates the source code's structure into an \astr  as Prolog facts.
To extract the \astr from the source code the converter uses Clang, a C/C++ front end for \llvm\footnote{Low Level Virtual Machine: http://llvm.org}.
The converter consists of about 190 lines of C code, and translates every node from the \astr into Prolog facts with the same structure, independent of its type.
Using the same fact structure eases processing the full \astr without paying attention to the types.
The resulting facts (\astr nodes) have the following structure:
\begin{quote}
\begin{lstlisting}[language=prolog]
node(File, Ast_Order, Id, Par_Id, Type, Src_Loc, Params).
\end{lstlisting}
\end{quote}

\noindent
For every C/C++ source file, Clang creates a separate \astr.
The node IDs are unique within a single \astr.
Our verification process, however, operates on the complete \astr of the full software, thus, it merges all the individual \astrs.
To keep the nodes unique, the converter extends them by the name (path) of the source file on which a single \astr bases (\verb|File|, first argument of \verb|node/7|).
The second argument (\verb|Ast_Order|) defines the relative location of a node among its siblings.
It is an integer value starting with 0 among the siblings.
The argument \verb|Id| represents the identifier of the node within a source file's \astr, and \verb|Parent_Id| is the identifier of its parent node in the same \astr.
The fifth argument (\verb|Type|) names the node Type (e.g.\ \verb|IfStmt|, \verb|CXXMethodDecl|).
The argument \verb|Src_Loc| contains the location of the node within the source code given by its file name, the line and the column of the first character as well as its range.
However, not every node contains the complete information about its location.
There are some information (e.g.\ the source file name) that have to be obtained from the ancestor nodes.

Depending on its \verb|Type|, a node comes with specific parameters which differ in number and in data types.
However, the converter translates every node into the same fact structure.
This enables a consistent processing of the \astr.
Thus, the additional information is saved in the last element (\verb|Parameters|) as a Prolog list which contains the specific extra information.
For accessing the extra information, particular access rules are defined (see end of Section~\ref{sec node rules}).

\paragraph{Example Representation of a Source Code as an \astr}

\begin{figure*}[htp]
\centering
\includegraphics[width=.9\textwidth]{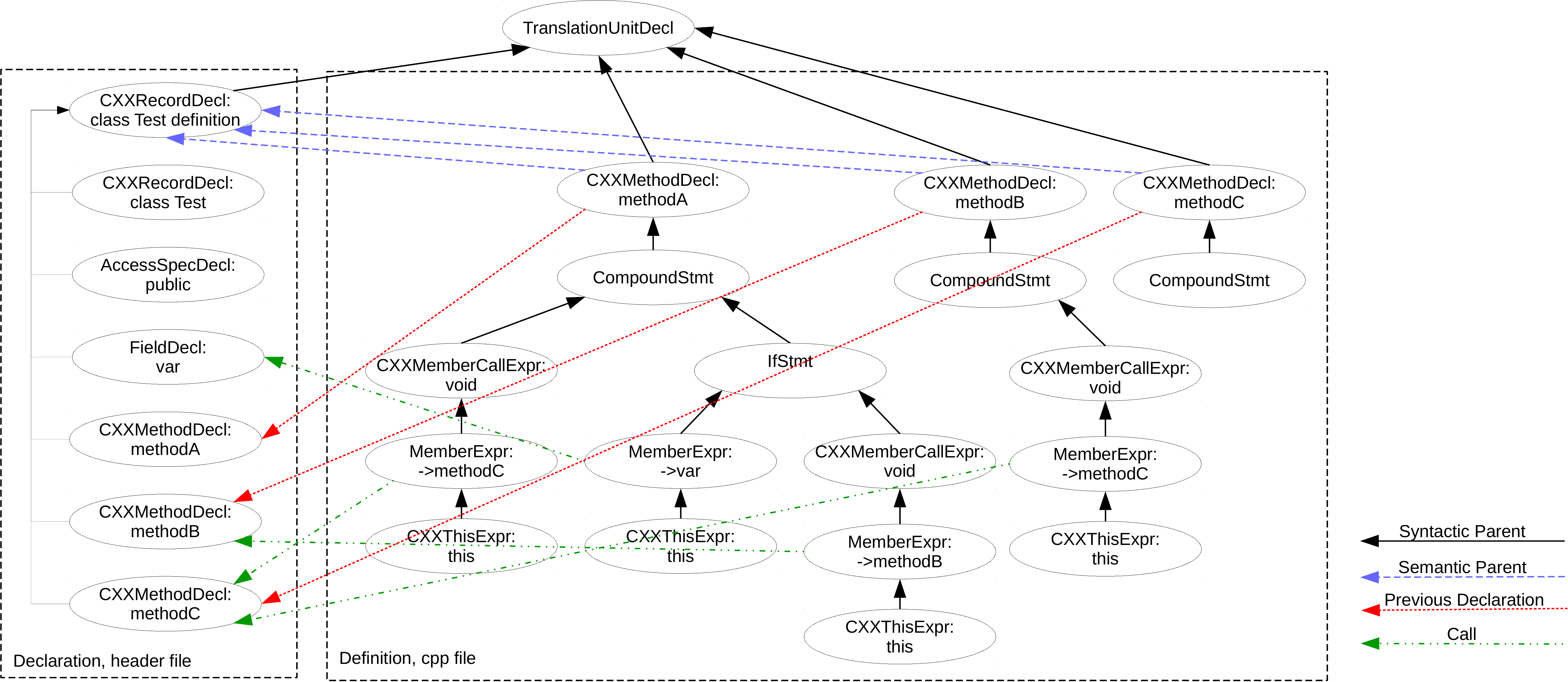}
\caption{Example of an \astr from Clang with Semantic References}
\label{fig ex asg split}
\end{figure*}

Apart from the syntactical structure of the source code, Clang additionally includes references that describe semantic links between nodes of the \astr.
Figure~\ref{fig ex asg split} shows an \astr for the file \verb|test.cpp| from the following source code:

\begin{quote}
\begin{lstlisting}[language=c++]
// File: test.h
class Test {   
public:
  bool var;
  void methodA();
  void methodB();
  void methodC();
};

// File: test.cpp
#include "test.h"
void Test::methodA() {
  methodC();
  if(var) methodB();
}
void Test::methodB() {
  methodC();
}
void Test::methodC() { }
\end{lstlisting}
\end{quote}

\noindent
The type of the root node is always \verb|TranslationUnitDecl|.
Due to \verb|#include| statements in the source file, Clang adds nodes from additional files, usually header files, to the \astr.
In the example, the \astr contains nodes from a source and a header file; Figure~\ref{fig ex asg split} separates the nodes by their origin (dashed boxes).
The header file declares the class \verb|Test| (node \verb|CXXRecordDecl|\footnote{The prefix CXX in the types of the nodes stands for C++.}), the member variable \verb|var| (node \verb|FieldDecl|) and the three methods \verb|methodA|, \verb|methodB|, and \verb|methodC| (nodes \verb|CXXMethodDecl|).
The implementations (definitions) of the methods are written in the source file, and not the header file.
Therefore, the method declaration nodes \verb|CXXMethodDecl| from the header file lack syntactic children (black edges).
Instead, they have semantic children (blue edges) that describe their implementations (definitions).
The implementations (definitions) of the methods, though, are syntactical children of the root node (\verb|TranslationUnitDecl|).

The body of a method is always enclosed in curly braces; the \verb|CompoundStmt| stands for this enclosed compound of statements.
Consequently, the children of \verb|CompoundStmt| describe the method's body.
A method call is defined by three subsequent nodes; \verb|CXXMemberCallExpr| specifies the return type, \verb|MemberExpr| describes the member name (method or variable), and \verb|CXXThisExpr| defines the variable (object) on which the call or access is executed.
Additionally, Clang adds a reference to the declaration of the called method to the \verb|MemberExpr| nodes (green edges).
Moreover, if a method is already declared before, Clang adds a reference from the \verb|CxxMethodDecl| to the previous declaration node (also \verb|CxxMethodDecl|, red edges).
The \verb|methodA| calls one or two other methods; \verb|methodC| is always called whereas the call of \verb|methodB| depends on the evaluation result of the if statement.
The left child of the node \verb|IfStmt| defines the condition part, and the right child defines the \emph{then part}.
Thus, the call of \verb|methodB| is only executed if the value of \verb|var| is \verb|true|.

Regarding the extra (colored) edges, the resulting structure, actually, is not a tree; instead, it is a directed acyclic graph (\textsc{Dag}).
As mentioned above, the process in the first stage transforms the \astr from Clang into Prolog facts. Here are some example nodes:
\begin{quote}
\begin{lstlisting}[language=prolog]
% node(+File, +AST_Order, +Id, +Par_Id, +Type, +Src_Loc, +Params) <-
node( 'src/test.cpp', 4, '0x2b8e2a0', '0x2b8d8f0', 'CXXRecordDecl',
   '<data/src/test.cpp:1:1, line:13:1>',
   ['class', 'Test', 'definition'] ).
node( 'src/test.cpp', 7, '0x2b8e4c0', '0x2b8e2a0', 'CXXMethodDecl',
   '<line:3:3, line:6:3>',
   ['methodA', 'void (void)'] ).
\end{lstlisting}
\end{quote}

\paragraph{Prolog Rules for Accessing \astr Structures}
\label{sec node rules}

For an easy access to the \astr we define several rules.
Two predicates are used for accessing syntactic relations; \verb|edge/3| represents the syntactic edges of the \astr, and \verb|transitive/3| determines ancestors or descendants of nodes.
Additionally, we define particular predicates for accessing the different node types.
As mentioned in a previous paragraph, different types of nodes include different extra information in the list \verb|Parameters|.
The rules extract the extra information and provide them as arguments of the predicate.
The following example shows the rules for \verb|edge/3|, \verb|transitive/3|, \verb|if_stmt/3|, and \verb|member_expr/5|.
\begin{quote}
\begin{lstlisting}[language=prolog]
% edge(-File, -Parent, -Child) <-
edge(File, Parent, Child) :-
   node(File, _, Child, Parent, _, _).

% transitive(-File, -Ancestor, -Child) <-
transitive(File, Ancestor, Child) :-
   edge(File, Ancestor, Child).
transitive(File, Ancestor, Child) :-
   edge(File, Parent, Child),
   transitive(File, Ancestor, Parent).

% if_stmt(-File, -Id, -AST_Order) <-
if_stmt(File, Id, AST_Order) :-
   node(File, AST_Order, Id, _, 'IfStmt', _).

% member_expr(-File, -Id, -AST_Order, -Name, -Callee_Id) <-
member_expr(File, Id, AST_Order, Name, Callee_Id) :-
   node(File, AST_Order, Id, _, 'MemberExpr', Parameters),
   append(_, [Name, Callee_Id], Parameters).
\end{lstlisting}
\end{quote}
The predicate \verb|if_stmt/3| represents a node with the type \verb|IfStmt|.
The predicate \verb|member_expr/5| represents a node with the type \verb|MemberExpr|, which describes the access to a member variable or method.
The rule extracts the name and the \verb|ID| of the called method (\verb|Callee_Id|) from the list \verb|Parameters|.

\section{Call Graph in Prolog}
\label{sec cg}

The \astr, created by Clang, provides references from calling methods to the called methods, see Section~\ref{sec ast}.
Using this information we generate a graph that only contains information about method calls.
We name the graph \emph{call graph} (\cg), which can be expressed as
\begin{equation*}
CG=\left(G_{CG}, I_{CG}, D_{CG}\right),
\end{equation*}
with
\begin{itemize}
\item $G_{CG}=\left(V_{CG}, E_{CG}\right)$ defines the graph structure with a set of nodes $V_{CG}$ and a set of edges $E_{CG}$.
\item $I_{CG}=\left(N_{CG}, L_{CG}, W_{CG}, n_{CG}, l_{CG}, w_{CG}\right)$ defines additional information for the nodes and edges; $N_{CG}$ is a set of method names, $L_{CG}$ is a set of locations within the source code, $W_{CG}$ is a set of variable names; $n_{CG}$ is a labeling function $n_{CG}:V_{CG}\rightarrow N_{CG}$, $l_{CG}$ is a labeling function $l_{CG}:E_{CG}\rightarrow L_{CG}$, and $w_{CG}$ is a labeling function $w_{CG}:E_{CG}\rightarrow W_{CG}$
\item $D_{CG}=\left(C_{CG}, c_{CG}\right)$ defines the nodes' dependence on conditions (control structures); $C_{CG}$ is a set of condition IDs, and $c_{CG}$ is a labeling function $c_{CG}:E_{CG}\rightarrow \left(C_{CG}\times\mathbb{N}\right)^n$ with $n\in\mathbb{N}_0$.
\end{itemize}

\noindent
For the investigation of different program flows, we use information about control structures.
Either one or several nested control structures within method bodies decide whether a method call is skipped.
The transformation process labels the edges with a set of control structure IDs that affect the method call.
Some control structures like \verb|if| ensure an exclusive execution of their children.
For example, the control structure \verb|if| executes either the \emph{then part} or the \emph{else part}~--~never both parts at the same time.
To cover the exclusive execution, the labeling function $c_{CG}$ assigns the ID of the control structure as well as an identifier of the branching.
For example, the nodes of the \emph{then part} are categorized with the same ID of the \emph{if} node and the same branch ID \emph{0}; the nodes of the \emph{else part} have the same ID of the \verb|if|, but a different branching ID \emph{1}.

\begin{figure}
\centering
\includegraphics[width=.7\textwidth]{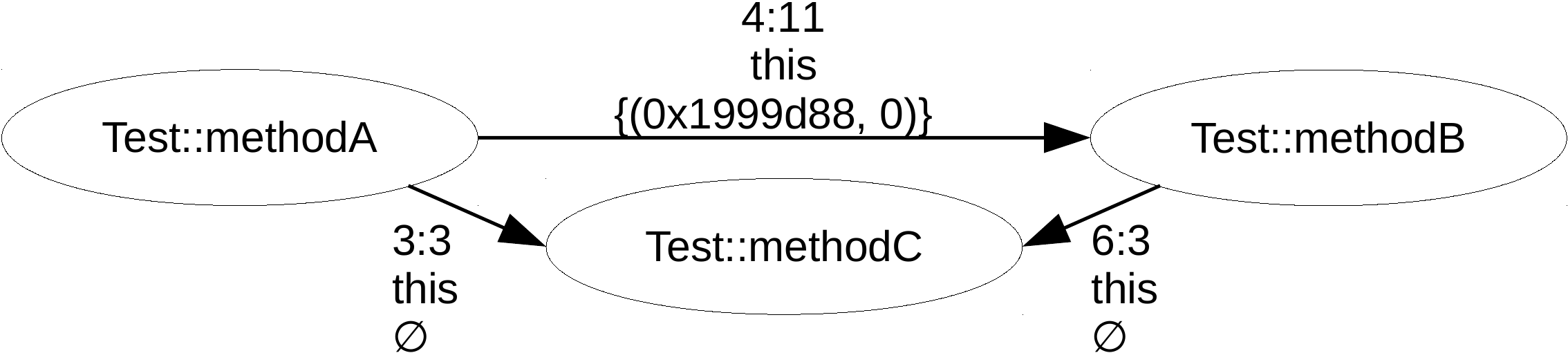}
\caption{Our example displayed as call a graph}
\label{test_cg}
\end{figure}

Figure~\ref{test_cg} shows the resulting \cg for the \astr from Figure~\ref{fig ex asg split}.
The nodes are labeled with the methods' names.
The edges are labeled with the source code location, the variable (object) name on which the call is invoked, and the set of the control structure and branch IDs  they depend on.

\paragraph{Detection Rules for Code Patterns of Methods}

Information about code structures like method declarations, their definitions, method calls etc. is stored within the \astr.
The information is not available in single nodes, but rather in subgraph structures of the \astr.
In Prolog, we define graph patterns especially for structures like method definition, method calls etc.
The following example shows rule that detects the declaration of a method.
\begin{quote}
\begin{lstlisting}[language=prolog]
% method_decl(-File, -Method_Id, -Full_Name) <-
method_decl(File, Method_Id, Full_Name) :-
   cxx_method_decl(File, Method_Id, Func_Name, _),
   transitive(File, Class_Id, Method_Id),
   cxx_record_decl(File, Class_Id, 'class', Class_Name),
   namespace(File, Method_Id, Namespace_Name),
   atomic_list_concat([Namespace_Name, '::', Func_Name], Full_Name).
\end{lstlisting}
\end{quote}

\noindent
Generally, the rule searches for a class definition which contains a method definition.
The rule seeks a pattern that consists of a node \verb|CXXMethodDecl| (predicate \verb|cxx_method_decl/4|) and an ancestor node \verb|CXXRecordDecl| (predicate \verb|cxx_record_decl/4|) with the record type \verb|'class'|.
The node for the method, \verb|CXXMethodDecl|, contains only the local name of the method, not the full one.
To obtain the full name, the rules of \verb|namespace/3| recursively traverse the \astr up and joins the appropriate scope names (e.g.\ from classes or namespaces).
After that, \verb|atomic_list_concat/2| merges the resulting prefix and the local name.

\paragraph{Detection Rules for Control Structures}

There are different types of structures that lead to conditional execution of operations.
On the one hand there are obvious keywords for control structures such as \verb|do|, \verb|for|, \verb|if|, \verb|while| and the ternary operator \verb|?|.
On the other hand there are structures which implicitly skip operations.
For example, an \emph{or} (\texttt{||}) executes the second operand only if the first one is false.
Otherwise the second operand is not executed at all  (short-circuit evaluation).

We define rules that identify control structures and determine the root nodes of the conditionally executed subtrees.
The following example shows a rule extracting an \verb|if| statement with a \emph{then part} and an \emph{else part}:
\begin{quote}
\begin{lstlisting}[language=prolog]
% if(-File, -Cond_Id, -Then_Id, -Else_Id) <-
if(File, Cond_Id, Then_Id, Else_Id) :-
   if_stmt(File, Id, _),
   node(File, 1, Cond_Id, _, _, _),
   edge(File, Id, Cond_Id),
   node(File, 2, Then_Id, _, _, _),
   edge(File, Id, Then_Id),
   node(File, 3, Else_Id, _, _, _),
   edge(File, Id, Else_Id).
\end{lstlisting}
\end{quote}

\noindent
The rule seeks a node \verb|IfStmt| (predicate \verb|if_stmt/3|) and its three children.
The three children are the root nodes of subtrees that stand for the different parts of the \verb|if|; the first subtree represents the \emph{condition part}, the second subtree represents the \emph{then part}, and the third subtree represents the \emph{else part}.
The order of the children defines their assignment to the parts of the \verb|if|; therefore, they are determined by their order 1 to 3 in \verb|node/6|.
We also defined a similar rule for \verb|if|s without an \emph{else part}, which is not shown in this paper.

\paragraph{Virtual Methods}
An important feature of the object-oriented programming paradigm is the inheritance of classes.
Subclasses can overwrite methods if the method is declared as \verb|virtual| in the superclass.
Due to inheritance, a pointer for a superclass can also point to a corresponding subclass.
Then, if an overwritten method is called, the implementation in the subclass is called instead of the one in the superclass.
As mentioned in Section~\ref{sec ast}, nodes that represent method calls refer to the declaration of the called method.
However, it always points to the method of the pointer's type, not to overwritten methods.
Thus, we generate multiple \cgs that call overwritten methods instead.

\section{Building Execution Sequence Trees in Prolog}
\label{sec est}
The verification process investigates consecutive method calls that are executed after a given method call.
A tree structure, which we name Execution Sequence Tree (\est), represents the subsequent method calls.
To create an \est, the generator preforms a depth-first search on the \cg.
Beginning from the given method call and regarding the program flow, the subsequent method calls (successor nodes) are added sequentially as a list to the \cg.
After traversing the successor nodes, the generator visits the parent nodes of the given method call.
The parent nodes represent calling methods and, therefore, a return back to the calling methods.
Each parent leads to a branching in the \est, which represent different program flows.

\paragraph{Definition of Execution Sequence Trees}
The tree structure \est can be expressed as
\begin{equation*}
EST = \left(G_{EST}, I_{EST}, D_{EST}\right),
\end{equation*}
where
\begin{itemize}
\item $G_{EST}=\left(V_{EST}, E_{EST}\right)$ defines the graph structure with a set of nodes $V_{EST}$ and a set of edges $E_{EST}$,
\item $I_{EST}=\left(N_{EST}, L_{EST}, W_{EST}, n_{EST}, l_{EST}, w_{EST}, t_{EST}\right)$ defines additional information for nodes and for edges; $N_{EST}$ is a set of method names, $L_{EST}$ is a set of locations within the source code, $W_{EST}$ is a set of variable names; $n_{EST}$ is a labeling function $n_{EST}:V_{EST}\rightarrow N_{EST}$, $l_{EST}$ is a labeling function $l_{EST}:V_{EST}\rightarrow L_{EST}$, $w_{EST}$ is a labeling function $w_{EST}:V_{EST}\rightarrow W_{EST}$, and $t_{EST}$ is a labeling function $t_{EST}:V_{EST}\rightarrow \{c, p, r\}$ for the set of call types: child ($c$), parent ($p$), and root ($r$),
\item $D_{EST}=\left(C_{EST}, c_{EST}\right)$ defines which nodes depend on conditions (control structures); $C_{EST}$ is a set of conditions IDs and $c_{EST}$ is a labeling function $c_{EST}:E_{EST}\rightarrow \left(C_{EST}\times\mathbb{N}\right)^n$ with $n\in\mathbb{N}_0$.
\end{itemize}

\noindent
An \est is specifically created for a given method call (edge within the \cg).
At first, the calling method is used as root node including the information about the call, i.e.\ the location in the source code and the variable (object) name.
The node becomes the active branch of the \est.
According to the program flow, traversing rules add subsequent method calls to the active branch with regard to the following guidelines:
\begin{description}
\item[Children]
  Coming from the entry edge, the current node is added to the active branch of the \est including the information of the entry edge.
  After that, all successors are traversed in a depth-first search, whereby siblings are ordered by their location within the source code.
\item[Parents]
  After processing all children or detecting a return node, the traversing rules proceed to the parent nodes.
  They add each parent node to the last node of the active branch of the \est.
  As a result, the traversing rules create new branchings which then are separately processed (becoming active), beginning with the children.
\item[Recursion]
  To avoid infinite loops in the traversing process, recursions are handled specially.
  If there is a recursion (direct or mutual method calls) within the \cg, then only one complete run is added to the active branch of the \est.
  After that, the depth-first search is aborted and backtracked to the next node after the entrance into the recursion.
  However, the information about which methods are called in the recursion remains in the \est due to the one complete run that was added.
\end{description}

\noindent
In the \cg, information about conditions are assigned to the edges.
To make the information also available in the \est, the transformation process adds it to the target node.
Additionally, the information is propagated for all successors from the \cg and added to the nodes of the \est appropriately.
Thus, the information about conditional execution of method calls is available for every node in the \est.

\begin{figure*}[htp]
\centering
\begin{subfigure}[b]{.5\textwidth}
  \centering
  \includegraphics[width=\textwidth]{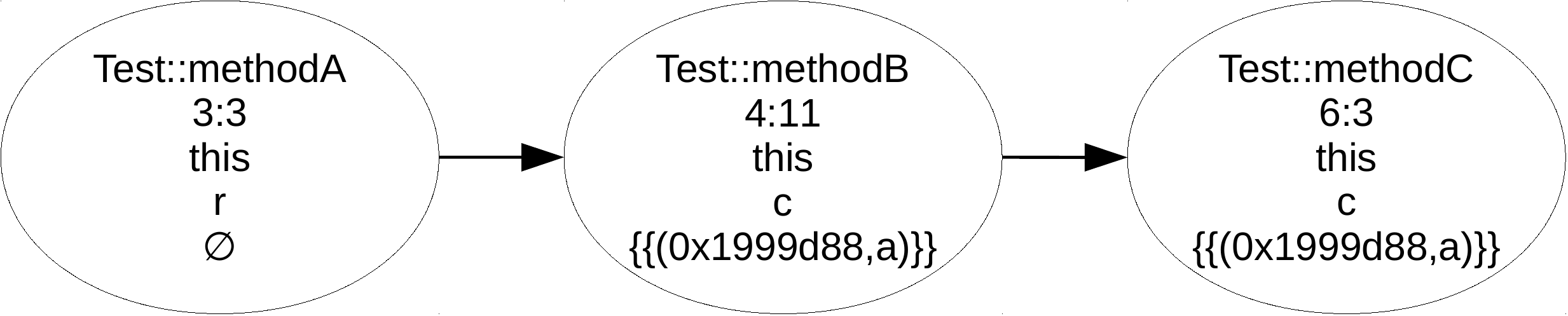}
  \subcaption{Full \est for methodA $\rightarrow$ methodC call}
\end{subfigure}\hspace{.05\textwidth}
\begin{subfigure}[b]{.4\textwidth}
  \centering
  \includegraphics[width=.8\textwidth]{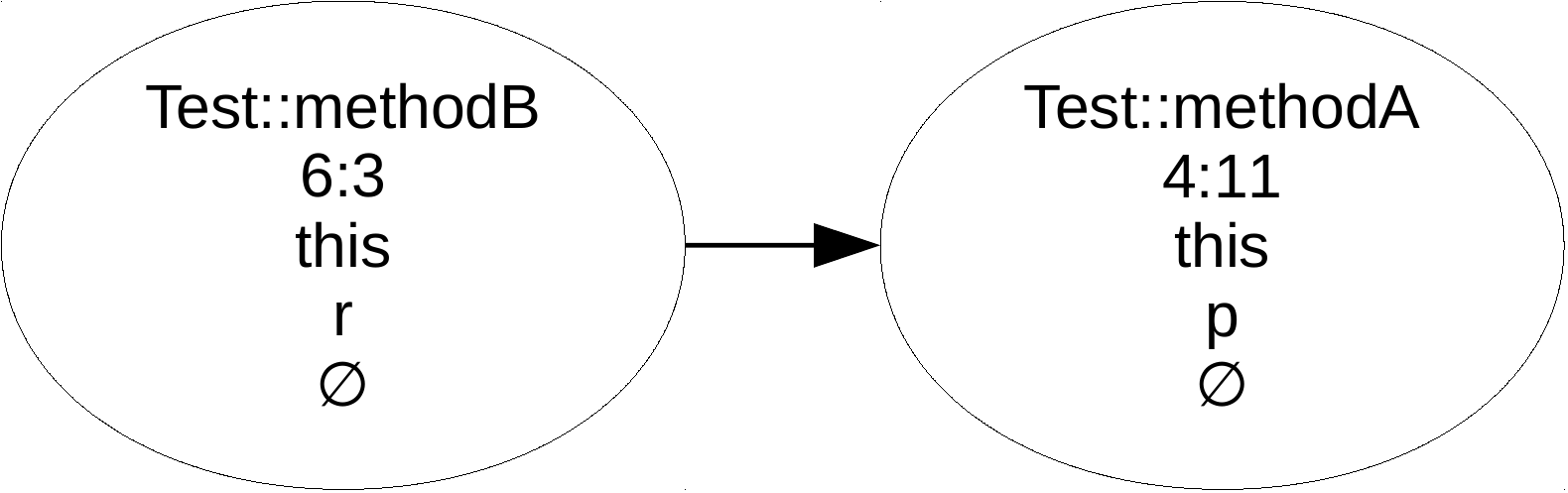}
  \subcaption{Full \est for methodB $\rightarrow$ methodC call}
\end{subfigure}
\caption{Execution Sequence Trees of the (Simple) Running Semaphore Example}
\label{fig ex sema est}
\end{figure*}

Keeping our example from Figure~\ref{test_cg}, Figure~\ref{fig ex sema est} shows two \ests which base on different calls of \verb|methodC|.
Both resulting \ests are lists rather than dendritic tree structures, because of the simple running example.
The first EST (Subfigure~(a)) bases on the method call \verb|methodA|$\rightarrow$\verb|methodC|.
Consequently, \verb|methodA| becomes the root node.
The depth-first search traverses all children that are called after the base call \verb|methodA|$\rightarrow$\verb|methodC|.
The location in the source code of \verb|methodA|$\rightarrow$\verb|methodB| (4:11) is greater than the base call \verb|methodA|$\rightarrow$\verb|methodC| (3:3).
Therefore, \verb|methodB| becomes the successor node in the \est.
After that, the process adds \verb|methodC| to the \est because \verb|methodB|$\rightarrow$\verb|methodC| is the next subsequent method call within \verb|methodB|.
Then, no other subsequent method calls are left to add.

The second \est (Subfigure~(b)) bases on the method call \verb|methodB|$\rightarrow$\verb|methodC|.
Consequently, the method \verb|methodB| becomes the root node.
There are no children in the \cg for \verb|methodB|, therefore, it proceeds to the parent nodes and adds its parent \verb|methodA| to the \est following the parent edge.
The process does not add any child nodes because the entry edge (\verb|4:11|) is underneath the other method call \verb|methodA|$\rightarrow$\verb|methodC| (\verb|7:3|) in the source code.
Thus, there are no other nodes to add.

\label{sec est_node}
The generation process creates a set of term structures \verb|est_node/4| which defines the a full \est.
Terms of \verb|est_node/4| have the following structure:
\begin{lstlisting}
   est_node(Id, Location, Variable, Type)
\end{lstlisting}

\noindent
The first parameter \verb|Id| defines the identifier for the node within the \est.
Unlike in \cgs, a method call can exist more than once within the \est, therefore, using the method name is not enough for a unique identification.
Instead, the full path of method calls from the current node to the root node is used as the identifier.
This enables additionally an easy access of previous method calls.
The second and third parameters \verb|Location| and \verb|Variable| are the same values as from the \cg.
The last parameter \verb|Type| is either \verb|r| (root), \verb|c| (child) or \verb|p| (parent).

\paragraph{Bringing Non-Determinism to the Execution Sequence Tree}
\label{sec nest}

\begin{figure*}[h]
\centering
\begin{subfigure}[b]{0.4\textwidth}
  \centering
  \includegraphics[width=\textwidth]{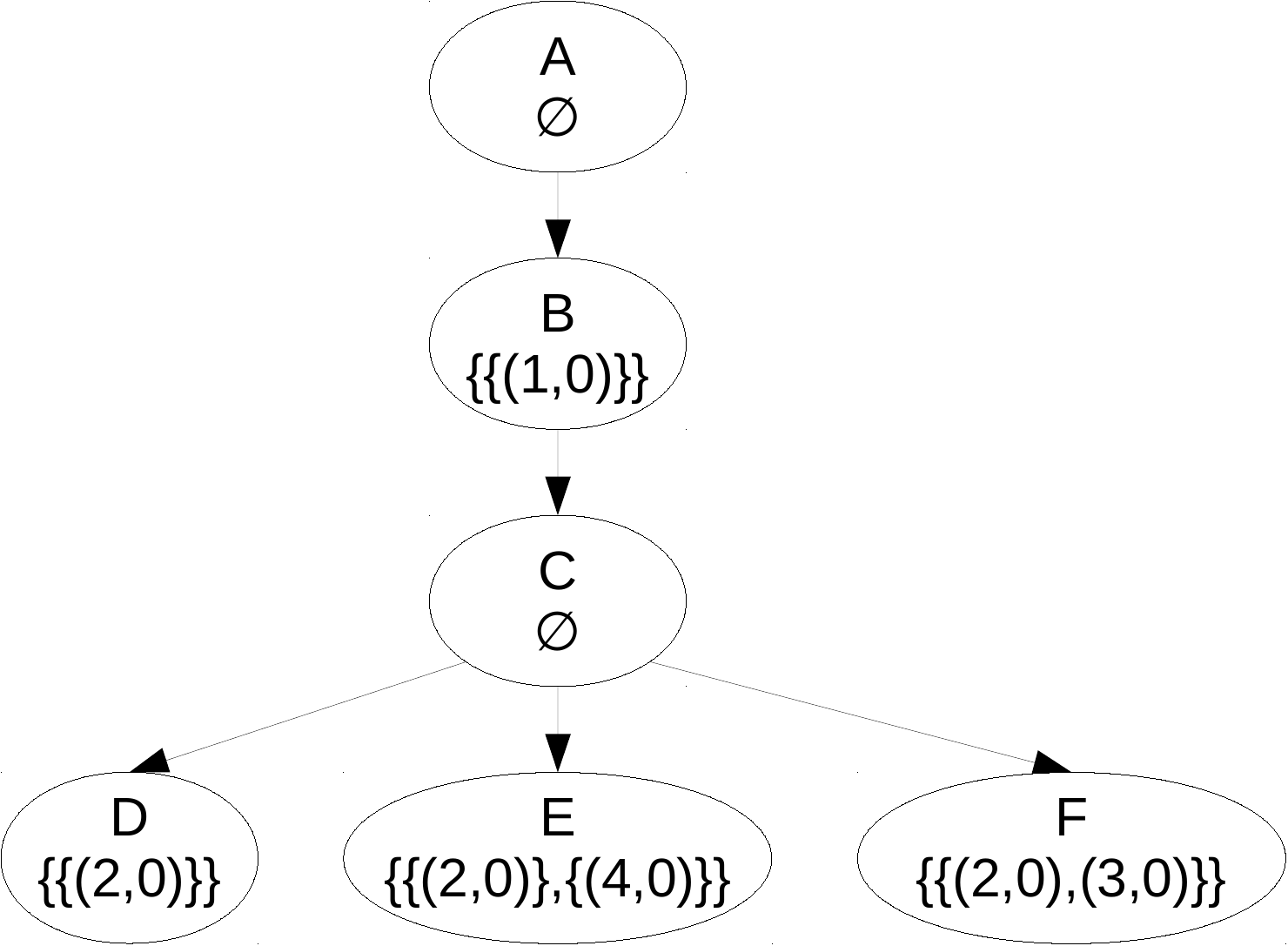}
  \subcaption{Example for a full \est}
\end{subfigure}
\begin{subfigure}[b]{0.4\textwidth}
  \centering
  \includegraphics[width=.2\textwidth]{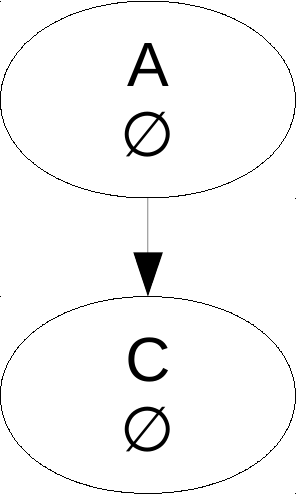}
  \subcaption{\nest without non-deterministic nodes}
\end{subfigure}\vspace{.5cm}
\begin{subfigure}[b]{0.4\textwidth}
  \centering
  \includegraphics[width=.575\textwidth]{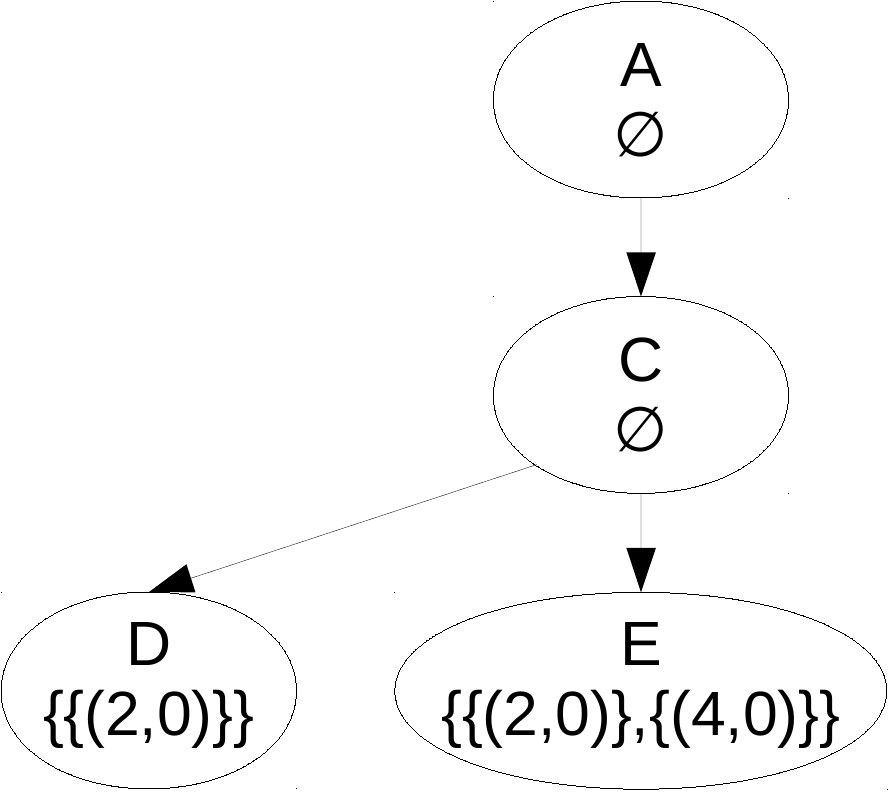}
  \subcaption{\nest with nodes for \{(2,0)\}}
\end{subfigure}
\begin{subfigure}[b]{0.4\textwidth}
  \centering
  \includegraphics[width=\textwidth]{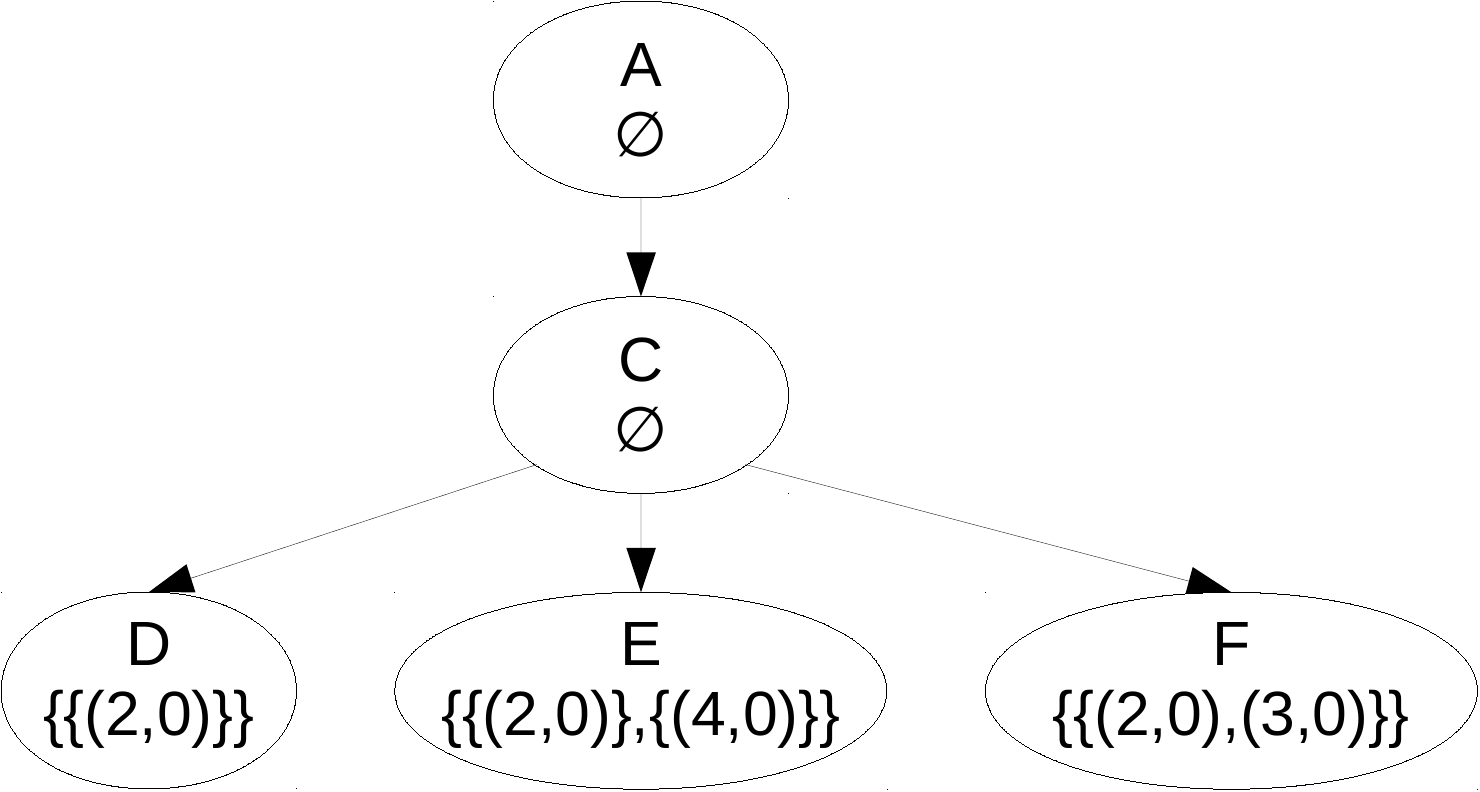}
  \subcaption{\nest with nodes for \{(2,0),(3,0)\}}
\end{subfigure}
\caption{Some examples for \nests for a given full \est}
\label{fig nest ex}
\end{figure*}

Control structures affect the execution of the software; some method calls are only executed in certain program flows.
To investigate different program flows for programming errors, non-deterministic Prolog rules create variants (subtrees) of \ests, each representing a program flow.
Due to the non-determinism in the creation process, we name a resulting variant of the \est \emph{non-deterministic execution sequence tree} (\nest).
A \nest for a given \est is identified by the set $I\subseteq C_{EST}\times\mathbb{N}$ of valid (true) control structures and their branch IDs (e.g.\ \emph{then} or \emph{else part}).
The set of nodes of a \nest $V_{NEST}\subseteq V_{EST}$ is the union of the two sets $V_{Det}$ and $V_{Nondet}$ with
\begin{itemize}
\item $V_{Det}=\left\{v\in V_{EST}\mid c_{EST}(v)=\emptyset\right\}$ is the set of nodes which are present in every variation of \nests (deterministic), and
\item $V_{Nondet}=\left\{v\in V_{EST}\mid c_{EST}(v)\subseteq I\right\}$ is the set of nodes which dependent on control structures. Therefore, it differs for every \nest variant.
\end{itemize}

\noindent
The edges of a \nest $E_{NEST}\subseteq E_{EST}$ are reduced according to the nodes.
Only edges whose both nodes are part of the \nest become an edge of the \nest: $\left(e=(v_1,v_2)\in E_{EST}\land v_1,v_2\in V_{NEST}\right)\Rightarrow e\in E_{NEST}$.
Additionally, if nodes are left out, their incoming edges are propagated as outgoing edges:

$\left(v_0,v_2\in V_{NEST}\land v_1\notin V_{NEST}\land e_{01}=(v_0,v_1)\in E_{EST}\land e_{12}=(v_1,v_2)\in E_{EST}\right)\Rightarrow (v_0,v_2)\in E_{NEST}$

\noindent
Accordingly, the same is true for more than one node left out in a row, as well.
Figure~\ref{fig nest ex} shows a complete \est and three variations of \nests.
For reasons of comprehensibility only the method names and the set of condition IDs are printed; other labellings are omitted.
The full \est in Subfigure~(a) contains nodes with varying control structure dependencies.
The nodes~\verb|B| and \verb|D| are valid for single conditions (\verb|(1,0)| resp. \verb|(2,0)|).
The node~\verb|E| depends on the two single conditions \verb|(2,0)| and \verb|(4,0)|.
If one of the control structures is valid, node~\verb|E| becomes part of the \nest.
The node~\verb|F| is only included in a \nest if both conditions, \verb|(2,0)| and \verb|(3,0)|, are valid.

Subfigures~(b) to (d) show different variants of \nests.
The \nest with no valid control structures is the least complex one.
It only contains nodes with empty sets of control structures, as Subfigure~(b) shows.
Every node with a set of control structures that are valid for a \nest (in $I$), becomes part of the \nest.
Subfigure~(c) shows the \nest with one valid control structure: \verb|(2,0)|.
The Node~\verb|F| contains a combined set \verb|{(2,0),(3,0)}|; however, the condition \verb|(3,0)| is invalid, therefore, the node is left out.
For \nest of Subfigure~(d) both, \verb|(2,0)| and \verb|(3,0)| are valid.
Consequently, nodes with only \verb|(2,0)| (nodes~\verb|D| and \verb|E|) and nodes with combined conditions (node~\verb|F|) are included in the \nest.

\paragraph{Limitation of the Range of \ctl Rules}
Rules for \ctl check the properties of nodes in a tree.
The number of checked nodes before the actual result is determined depends on the rules, on the properties of the nodes and on the quantifiers.
The Table~\ref{tab ctl quant} gives an overview of the range of \ctl quantifiers. As soon as a success or a fail is determined, the \est creation stops and ignores the rest of the tree.
\begin{table}[htp]
\begin{tabular}{r||l|l||l|l}
\multirow{2}{*}{\parbox[t]{.95cm}{Quan\-tifier}} & \multicolumn{2}{c||}{Success} & \multicolumn{2}{c}{Fail} \\
& Condition & Ignored Nodes & Condition & Ignored Nodes\\
\hline
\hline
AF & a \parbox[t]{4.3cm}{node per path successes} & \parbox[t]{2.6cm}{remaining nodes per path} & \parbox[t]{3.0cm}{all nodes in a path fail} & remaining paths\\
\hline
AG & all nodes success & none & one node fails & remaining nodes\\
\hline
AX & all next nodes success & none & a next node fails & \parbox[t]{2.6cm}{remaining next nodes per path}\\
\hline
AU & \parbox[t]{4.3cm}{1st stmt. successes until 2nd stmt. successes per path} & \parbox[t]{2.6cm}{remaining nodes per path} & \parbox[t]{3.0cm}{1st and 2nd stmt. fail} & remaining nodes\\
\hline
EF & one node successes & remaining nodes & \parbox[t]{3.0cm}{all nodes in all paths fail} & none\\
\hline
EG & \parbox[t]{4.3cm}{all nodes for a path success} & remaining paths & \parbox[t]{3.0cm}{a node of the last path fails} & \parbox[t]{2.55cm}{remaining nodes of the last path}\\
\hline
EX & \parbox[t]{4.3cm}{a next node of a path successes} & remaining nodes & \parbox[t]{3.0cm}{next nodes in all paths fail} & none\\
\end{tabular}
\caption{Maximum Range of \ctl Rules (stmt. = statement)}
\label{tab ctl quant}
\end{table}

%\paragraph{Generating Non-deterministic Execution Sequence Trees in Prolog}
Prolog rules traverse the \cg and construct the \nests, simultaneously applying the user-defined \ctl propositions.
If the \ctl proposition success prematurely, the creation process aborts and proceeds with the next \nest.
However, if the \ctl proposition fail, the process aborts and the user is informed about the problematic \nest.
Consequently, the creation process for \nests neglects control structures that are only used in ignored subtrees; thus, the number of created \nests decreases.

\section{Formulating Verification Goals for Semaphores}
\label{sec case study}
In Section~\ref{sec verification goals} we presented semaphores, which are provided as classes in the operating system \rodos.
They are used by invoking their methods \verb|enter| and \verb|leave|.
We only analyze the static source code, therefore, there are limitations regarding variable values.
For example, we cannot detect whether two pointers refer to the same object, because we do not have access to the actual values.
However, semaphores in \rodos are usually accessed by using their object variable and seldom by using pointers.
Therefore, the static source code analysis is sufficient for our semaphore investigations.

\paragraph{Semaphore Entered but Never Left}
The paragraph about verification goals in Section~\ref{sec verification goals} illustrates an incorrect usage of semaphores: entering a semaphore without leaving it. We generate a \cg and consecutively derive an \est, based on the method call \verb|Test::methodE|$\rightarrow$\verb|Semaphore::enter|.
Then, we apply a \ctl proposition on the \est:
\begin{equation*}
  AF\left(n = Semaphore::leave \land w=w_r\right)
\end{equation*}
On every branch of the \est ($A$), finally ($F$), the statement should be fulfilled: a node ought to have the method name $n$ = \verb|Semaphore::leave|.
Additionally, the variable (object) name $w$ must be the same as the one from the root node ($w_r$), to assure that the same semaphore is investigated.
In Prolog, we define \ctl statements as binary predicates.
The first argument is a term structure \verb|est_node/4| (see Section~\ref{sec est_node}) of the root node in the \est, the second one is a term structure \verb|est_node/4| of the currently investigated node of the \est:
\begin{quote}
\begin{lstlisting}[language=prolog]
% sema_leave(+Root, +Node) <-
sema_leave(Root, Node) :-
   Root = est_node(_, _, Var, _),
   Node = est_node(['Semaphore::leave'| _], _, Var, _).
\end{lstlisting}
\end{quote}
The variable name (\verb|Var|) has to be the same.
Additionally, the name of the called method must be \verb|Semaphore::leave|.
This is done by peaking on the top element of the Id that represents the path of method names from the current node to the root node of the \est.

\paragraph{Semaphore Entered Several Times before Left}
Secondly, we check whether a semaphore is entered twice or more often before it is left.
An example clarifies the issue:
\begin{quote}
\begin{lstlisting}[language=c++]
void SemaTest::methodH() {
  sema.enter();
  methodB();
  sema.enter();
  methodB();
  sema.leave();
}
\end{lstlisting}
\end{quote}
The semaphore \verb|sema| is entered before and after the first call of \verb|methodB|, but left only once, after the second call.
This programming error leads to an infinite blocking of the program at the second call of \verb|sema.enter()|.
To find this issue in the source code, we describe the correct usage in \ctl:
\begin{equation*}
A(\lnot(n=Semaphore::enter \land w=w_r)~U~(n=Semaphore::leave \land w=w_r))
\end{equation*}
This \ctl proposition consists of two separate statement; for all branches ($A$) the first one must be true until ($U$) the second one becomes true.
The first statement expresses that the investigated node must \emph{not} have the method name $n$ = \verb|Semaphore::enter| with the same variable (object) name $w$ as the one of the root node ($w_r$).
The second statement expresses that the investigated node \emph{must} have the method name $n$ = \verb|Semaphore::leave| on the same variable (object) name $w$ as the one of the root node ($w_r$).
That means, that a semaphore should not be entered before it is left.

Above, we had already written the second statement in Prolog; the first statement in Prolog follows:
\begin{quote}
\begin{lstlisting}[language=prolog]
% sema_not_enter(+Root, +Node) <-
sema_not_enter(Root, Node) :-
   Root = est_node(_, _, _, Var_1, _),
   Node = est_node(_, [Name_2| _], _, Var_2, _),
   ( Var_1 \= Var_2
   ; Name_2 \= 'RODOS::Semaphore::enter' ).
\end{lstlisting}
\end{quote}
This rule successes if the variable (object) names or the method names of the root and the current node differ.

\section{Other Approaches Using Logic Programming for Verification}
\label{sec related work}

There exist several approaches which introduce logic programming
into the static source code analysis of software.
A brief comparison between logic-based infrastructures concerning detection
and extraction is given by Kniesel, Hannemann and Rho \cite{Kni:07},
who compare different frameworks that enable software analysis and
manipulation by an object-oriented program representation.
The comparison addresses efficiency and scalability,
but also further criteria such as expressiveness, turnaround
and availability.
The comparison also includes the formulation of a design pattern detection.
They compare two frameworks for the analysis of Java source code:
JQuery and their own approach JTransformer/CTC.
CodeQuest was selected as a reference for performance and scalability.
% As a result of the comparison, JTransformer seems to be a good choice.
However, JTransformer is limited to process Java source code.
There exists an extension to JTransformer, named
StarTransformer\footnote{%
https://sewiki.iai.uni-bonn.de/research/jtransformer/api/meta/startransformer},
which enables the analysis of other languages as well, but no application for
C/C++ could be found yet.
JTransformer is designed as a plug-in solely for the use with Eclipse\footnote{https://www.eclipse.org}, a popular integrated development environment (\ide) for Java.
But the development of embedded software often depends on different \ides or command line tools.

Consens and Mendelzon describe in various publications \cite{ConMenAlb:90, ConMend:93} their query language named GraphLog.
It operates on data represented as graph structures, even the queries are described as graph patterns.
To evaluate a query the graph pattern is searched within the database graph.
Data, queries and results can be visualized.
The authors argue that many databases can be viewed as graphs.
As a case study, they use GraphLog to analyze software structures in \cite{ConMendRym:92}.
Consens et al. examine the package structure of a complex Windows application in order to remove cycles within the package dependency.
However, the data structure for their software analysis is not the full \astr.
Therefore, this approach is highly specialized and difficult to generalize.

Ciraci describes in his dissertation \cite{Cir:09} the graph-based verification
of static program constraints \cite{CirBroAks:10}.
In the first step, the original source code is automatically converted into a
intermediate representation which is called \scml (Source Code Modeling
Language).
\scml is an attributed graph representation of the source code.
The \scml expressions can be imported directly from different
languages~-- there are already converters for Java and C/C++.
The constraints for the source code are
described as graph transformation rules which also written in \scml.
To notify the user for violated constraints, information nodes are inserted locally into the original \scml.
The user can write Prolog rules to further investigate the information nodes.
However, the program flow is not covered in Ciraci's analysis.

Crew \cite{Cre:97} created a \astlog, a Prolog-like programming language, to specifically process \astrs. 
\astlog does not save a given \astr into an internal database, but references the elements of the \astr directly.
A difference to a fact base in Prolog is the Current Object which only is implicitly available to rules.
The rules are evaluated in the context of the Current Object; similar to a visitor in the visitor pattern.
Even \astlog is Prolog-like, compatibility is not given.

Centaur \cite{BorCleEtAl:88} is a generic interactive debugging system.
The input is a formal specification of the syntax and semantics of the used programming
language.
The specifications are described in Prolog which are used
to convert the given source code into the internal data format (Virtual Tree Processor).
Albeit Centaur provides an interactive investigation process, the automatic pattern recognition is not its main feature and therefore hard to modify.

Ballance, Graham and Van De Vanter present Pan \cite{BalGraVant:90}, an integrated development
environment, which allows to analyze source code loaded by its editor.
They use logic programming in conjunction with Logical Constraint
Grammars differently from Prolog.
For instance, the logic database is partitioned into several collections and code in Pan is interpreted separated from data which makes it less flexible compared to homoiconic Prolog.

Wahler, Seipel, Von Gudenberg and Fischer propose an \xml structure for the 
representation of abstract syntax trees \cite{WahEtAl:04}.
On that structure, they apply an algorithm inspired from data mining
for searching clones within the source code by finding frequent itemsets in the \xml.
In case studies they run the analysis on the Java Development Kit and the 
Dislog Development Kit.
Additionally to the program analysis, the tool Squash of
Boehm, Seipel, Sickmann and Wetzka can be used for designing,
analyzing and refactoring relational database applications
\cite{BoeSeiEtAl:07}.
In their approach, schema definitions and queries from \sql are
mapped to an \xml representation called SquashML.
Like an \astr the \sql representation in SquashML is a tree structure.
By using the \xml query and transformation language FnQuery, they use Prolog rules to describe modifications of the structure of the relational database.
Additionally, Squash provides a visualization of the relations and join paths 
within \sql queries.

\section{Conclusions and Future Work}
\label{sec conclusion}
Using Prolog provides a flexible and concise definition of verification goals.
Furthermore, it enables the definition of domain-specific rules for validation.
For future work, we intend to demonstrate the versatility by further evaluations for the flash file system.
We aim at providing a verification that precisely addresses issues of flash file system, for example the retention of data in case of unpredicted failures.
Using Prolog's non-determinism enables the effortless investigation of different program flows.
Additionally, we are interested in further applications of Prolog's features like non-determinism for embedded software, beyond file systems.

The \est, which we introduced in Section~\ref{sec est}, allows the application of \ctl for method calls in the program flow to investigate programming errors.
By an early application of \ctl during the creation process of \nests, we already reduce the number of generated \nests.
To further decrease the number of \nests, we are going to introduce further techniques such as Symbolic Execution to our process.
Symbolic execution analyzes the domain of the values of variables, which reduces the non-determinism for control structures.

\bibliographystyle{eptcs}
\bibliography{paper}

\end{document}